# MAGNETIC STRUCTURAL EFFECT
# IN NONEQUILIBRIUM DEFECTIVE SOLIDS

F. Kh. CHIBIROVA

*Karpov Institute of Physical Chemistry, 105064 Moscow, Russia*
chibir@cc.nifhi.ac.ru

Scientific study of the effect of structural memory of nonequilibrium defective solids about the processing in magnetic field (the magnetic structural effect (MSE) was continued in this paper. The study was aimed to reveal the universal nature of the MSE, which was investigated in several new nonequilibrium defective solids. The results of investigation of the processing in the vortical magnetic field (PVMF) and its effect on the structure of the natural magnetite $Fe_3O_4$ and the $SnO_2$ films were presented. The methods of Mössbauer and X-ray spectroscopy were used. The PVMF reduction of a defectiveness of $Fe_3O_4$ structure in the magnetite was detected. The MSE was also observed in the Mössbauer spectra of diamagnetic tin oxide $SnO_2$ films after the PVMF. One of the possible explanations of the MSE was given in the paper.

*Keywords*: nonequilibrium defective solids, the processing in magnetic field, Mössbauer and X-ray spectroscopy.

**1. Introduction.**
In previous articles [1-3], it was shown, that in the Mössbauer spectra of defective oxides (HTSC-ceramics $YBa_2Cu_3O_{7-x}$ with $Fe^{57}$ impurity and cobalt oxide $Co_{3-x}O_4$ films), obtained after expose to the magnetic field, the effect of the structural memory (the magnetic structural effect (MSE) was observed.

For example, in [1] the Mössbauer absorption spectra of $YBa_2Cu_3O_{7-x}$ samples, synthesized from the initial oxides processed in the vortical magnetic field (H=0.1T), the MSE manifests itself in the increase of the Lamb-Mössbauer factor *f'* (about 20 %) in comparison with the "unprocessed" samples and in an unusually high oxygen concentration for a tetragonal phase (x~0,2). After processing of $Co_{3-x}O_4$ films [2, 3] in the constant magnetic field (H=3,5T) at temperatures higher and lower than Neel temperature $T_N$ ($T_N$=26K), both, the Lamb-Mössbauer factor *f'* and the relative contributions to a total spectrum of $Co_{3-x}O_4$ structure, changed in the Mössbauer emission spectra. In both cases the observed changes reflected the reduction of the defectiveness of structure. Thus, the MSE in the studied nonequilibrium oxides resulted in a closer to equilibrium state of the nonequilibrium oxide system.

In order to reveal the universal nature of the MSE, our present activity is directed both on the theoretical search of a complete chain of the events initiated by a magnetic field at the microscopic and macroscopic levels in the nonequilibrium defective solids and on the expansion of the list of nonequilibrium defective solids, in which the MSE was investigated.

The first step in creating of the general theory of the MSE was the construction of a microscopic model. The assumption was made that the nature of the MSE could be expressed as direct influence of a magnetic field on charged dot defects of crystal structure, for example, on cation and anion vacancies. In [4] the theoretical model of a low-energy electron trap (a charged dot defect) in a magnetic field was considered.

As the MSE showed itself as a change in the defectiveness of structure of nonequilibrium defective solids after the processing in a magnetic field, it was natural to expect a change of some physical and chemical properties, which depended on the state of charged dot defects in a solid structure. Therefore, the MSE can be observed experimentally not only directly at studying of the defectiveness of a solid structure, but also at studying of "secondary" magnetic effects, i.e. at studying of changes in some physical and chemical properties after the processing of solids in a magnetic field. Observation of the "secondary" MSE effects allows the number of investigated objects to be broaden significantly.

First, it refers to the investigation of plastic deformations in solids. Dislocations are the elementary carriers of plastic deformation. In numerous articles available on the magnetoplastic effect the movement of dislocations in a magnetic field are associated with a magnetic field interaction with charged dot defects (stoppers of dislocations) [5-7].

Literature data on different "secondary" magnetic effects are abundant. So, according to Speriosu et al [8], the sample of $\gamma$-$Fe_2O_3$, produced by oxidation of magnetite $Fe_3O_4$, after the first test experiment in magnetic field exhibited the experimental results typical for the initial $Fe_3O_4$. Many works [9-12] are also devoted both to the effect of various magnetic fields on the properties of semiconductor crystals and to the interpretation of these effects.

In this paper we continue the direct study of changes of the defectiveness of structure after a magnetic field processing of solids. The experimental results of the effect of processing on the structure of natural magnetite $Fe_3O_4$ and tin oxide $SnO_2$ film in the vortical magnetic field (PVMF) are submitted.

All studied oxide systems [1-3] obtained a magnetic structure. This fact caused certain doubts in the nature of the MSE. The research of the diamagnetic $SnO_2$ films showed, that the MSE is not associated with the magnetic nature of a processed material.

**2. Experimental.**
Samples were obtained from two different sets of natural $Fe_3O_4$ powders.

Samples of $SnO_2$ films were prepared by a magnetron deposition of metal tin $\beta$-Sn with further oxidation in air at $800^0C$ for 4-8 hours. Quartz was used as a substrate. The samples contained 1-2 mg of isotope $^{119}Sn$ per $sm^2$.

The magnetic vortical field with intensity H=0.1T and the frequency of 10 Hz was created in the device of vortical layer (DVL).

$^{57}Fe$ Mössbauer absorption spectra of $Fe_3O_4$ samples were recorded in the constant acceleration mode using a conventional spectrometer, equipped with a cryostat for work at low temperatures (less than 77K). $^{57}Co(Cr)$ was used as a source of $\gamma$-radiation. All values of chemical shifts were submitted vs. $\alpha$-iron. FeAl was used to detect the resonant counter with an absorber in a form of thin foil of alloy.

$^{119}Sn$ Mössbauer absorption spectra of $SnO_2$ films were recorded in the constant acceleration mode using a conventional spectrometer. The resonant counter of conversion electrons was used to register the Mössbauer spectra. The anode was used as a sample of $SnO_2$ film. Ca $^{119}SnO_3$ was used as a source of $\gamma$-radiation.

X-ray measurements were carried out on a DRON-3M diffractometer ($Cu_{K\alpha}$-radiation), equipped with a special device for film recording. The "Profit VZ" program was used for spectra adjustment. The "Powder" program was applied to calculate the parameters of a crystal cell.

**3. Results.**
A. Mössbauer spectra.

1. Magnetite $Fe_3O_4$.

Figure 1 and Figure 2 show the typical Mössbauer spectra recorded for the samples of magnetite $Fe_3O_4$ at room temperature (Fig.1) and at T=77K (Fig.2) prior to the PVMF and after it, correspondingly. The spectra of all investigated samples show several changes after the PVMF in comparison with the Mössbauer spectra of the initial samples.

The observed changes in complex spectra of $Fe_3O_4$ samples after the PVMF can be interpreted in the context of properties of the magnetite structure. In the stoichiometric magnetite, 16 tetrahedral sites of the $Fe_3O_4$ spinel structure are occupied with $Fe^{3+}$ ions (A sublattice) and 10 octahedral sites are occupied with $Fe^{3+}$ and $Fe^{2+}$ ions (B sublattice), correspondingly. At temperatures higher than T=119K, fast electron exchange in ionic pairs $Fe^{3+}$-$Fe^{2+}$ (B sublattice) takes place. Therefore in the Mössbauer spectrum of B sublattice only one sextet (instead of two sextets from $Fe^{3+}$ and $Fe^{2+}$ ions) is observed. Thus, the resulting spectrum of magnetite at T>119K exhibits a superposition of two magnetic sextets from A and B sublattices. At T<119K the resulting spectrum of magnetite becomes more complex due to the superposition of three magnetic sextets: one from $Fe^{3+}$ ions in A sublattice and two from $Fe^{3+}$ and $Fe^{2+}$ ions in B sublattice.

a) The Mössbauer spectra of $Fe_3O_4$ samples recorded at room temperature (T>119K).

In both Mössbauer spectra recorded for $Fe_3O_4$ samples at room temperature prior to the PVMF and after it (Fig.1), the sextet with higher hyperfine field $H_{hf}$=49,3T (chemical shift δ=0.30 mm/s) corresponds to A sublattice and the sextet with $H_{hf}$=46,0T (chemical shift δ=0.65 mm/s) corresponds to B sublattice. The spectra shown in Fig.1 (a) and Fig.1 (b) differ in an intensity of sextets lines and, consequently, in a ratio of the spectral areas of A and B sublattices.

Topsøe et al. [13] investigated in details the dependence between a degree of defectiveness of magnetite structure and a form of Mössbauer spectrum. They showed a high accuracy of coincidence of the Lamb-Mössbauer factor *f'* for both sublattices of magnetite. Therefore the degree of a deviation of magnetite structure from stoichiometry can be appreciated by a ratio S of spectral areas of two sextets of A and B sublattices: $S=S_B/S_A$ ($S_A$ – the spectral area of A sublattice sextet, $S_B$ – the spectral area of B sublattice sextet). In [13] the empirical formula *C*=(2-S)/(5S-6) for calculation of a concentration *C* of superstructural cation vacancies in magnetite was deduced.

We estimated the concentrations of superstructural cation vacancies *C* for the initial magnetite structure (S=1,64; Fig.1 (a)) and the processed sample structure (S=1,89; Fig.1 (b)). It turned out that concentration of the initial sample was *C*=0,16, while concentration of the processed sample was *C*=0,03.

Obviously *C* can be associated with the defectiveness of the solid structure, because dot defects make a considerable contribution to the total defectiveness. Therefore we can conclude that the PVMF reduces the defectiveness of the $Fe_3O_4$ samples.

b) Mössbauer spectra of $Fe_3O_4$ samples recorded at 77K (T<119K).

As indicated above, the Mössbauer spectra of $Fe_3O_4$ samples recorded at 77K prior to the PVMF and after it (Fig.2) show the superposition of three sextets. The spectra are not very informative because of their complexity, but they have a characteristic feature, which permits to evaluate concentration of $Fe^{2+}$ ions, associated with the stoichiometry of $Fe_3O_4$ in the velocity region of ~3 mm /s. Fig. 2 shows that the line of $Fe^{2+}$ ions is observed in both spectra in the velocity region of ~3 mm /s, but in the spectrum of the processed sample this line is more intensive, i.e. the concentration of $Fe^{2+}$ ions in this sample is higher. This result correlates well with the results obtained for spectra recorded at room temperature: the higher the stoichiometry of magnetite, the higher concentration of $Fe^{2+}$ ions in it.

All spectra of the magnetite samples after the PVMF have the same feature, i.e. the increase of the intensity of spectral lines. Such changes in the spectrum of the same magnetite sample can be explained only by the increase of the Lamb-Mössbauer factor *f'* for the magnetite sample after the PVMF. It is particularly visible in the magnetite spectra obtained at 77K (Fig.2).

2. $SnO_2$ films.

Figure 3 shows the typical Mössbauer spectra recorded for thin $SnO_2$ film samples at room temperature prior to the PVMF (Fig.3 (a)) and after it (Fig.3 (b)). Mössbauer spectra parameters are submitted in Table 1.

The spectrum in Fig. 3 (a) consists of two quadruple doublets: one relates to the basic $SnO_2$ phase and another relates to the defective surface SnO phase. After the PVMF, the second quadruple doublet disappears and only the quadruple doublet of $SnO_2$ phase remains in the spectrum in Fig.3 (b). The intensity of $SnO_2$ doublet increases almost on 20%. The increase of the intensity of $SnO_2$ doublet is relates to the increase of the Lamb-Mössbauer factor $f'$ for $SnO_2$ phase, as it was in the magnetite samples.

B. X-ray measurements.
The X-ray measurements were carried out for direct experimental observation of structural changes in $Fe_3O_4$ magnetite and thin $SnO_2$ films after the PVMF.

1. Magnetite $Fe_3O_4$
The calculated parameters $a$ and $d$ for $Fe_3O_4$ crystal cell and the cell volume $V$ of $Fe_3O_4$ samples measured prior to the PVMF and after it are presented in Table 2. The analysis of these results shows, that: (1) samples of the initial $Fe_3O_4$ are not stoichiometric, because the values for a lattice parameter $a$ (Table 2), calculated from X-ray spectra, differ from $a$=8,394 Å [JCPDS] for stoichiometric $Fe_3O_4$; (2) the $Fe_3O_4$ lattice parameter $a$ changes after the PVMF. In the samples processed in a magnetic field the volume $V$ of elementary cell decreases by 0,5 %. This agrees with the above conclusion about the reduction of concentration of cation vacancies in the sample of magnetite $Fe_3O_4$ after the PVMF.

2. $SnO_2$ films.
Table 3 presents the calculated parameters $a$ and $c$ of $SnO_2$ crystal cell for the samples of $SnO_2$ films measured prior to the PVMF and after it. According to these results, $SnO_2$ phase in the initial films is not stoichiometric, because the values of $a$ and $c$ differ from $a$=4,738Å and $c$=3,188Å [JCPDS] for stoichiometric $SnO_2$.

No SnO phase is found in the X-ray spectra of the $SnO_2$ film samples recorded prior to the PVMF and after it due to low concentration of this surface phase.

The conditions of X-ray recording of the thin film prevent obtaining highly accurate values of $a$ and $c$ constants for $SnO_2$ rutile lattice, thus hindering the evaluation of changes of these parameters after the PVMF. However, it is worth noting that after the PVMF the intensities of the spectral lines changed. The intensity of the (110) reflection increases remarkably, almost by 10%.

**4. Discussion.**
It is well known that in oxide systems a deviation from a stoichiometry is more often observed in surface layers. It was supposed that there exists a transitive layer, in which the transition from the broken order on the surface of a crystal to the order in the volume of a crystal takes place [14,15]. In existing oxides, formation of a transitive layer is caused by formation of defects with removal of several surrounding atoms or usually, with deeper structural reorganization and formation of new surface compounds.

A transitive layer is also formed on the surface of the magnetite crystallites. According to the stoichiometry of spinel structure of $Fe_3O_4$, there should be 33,33 % of $Fe^{2+}$ ions and 66,66 % of $Fe^{3+}$ ions. However, even in synthesized samples, let alone the natural magnetite, depending on the size of crystallites, the amount of $Fe^{2+}$ ions does not exceed 26 % for the size of 200nm and 31 % for the size of 1600nm [16]. These data show that the surface structure of magnetite crystallites is more defective and nonstoichiometric. Numerous researches [17, 18] showed that the surface of magnetite crystallites is enriched with the $Fe^{3+}$ ions, which form a surface $\gamma$-$Fe_2O_3$ phase, stabilizing a magnetite. The formation of $\gamma$-$Fe_2O_3$ on the magnetite surface can be formally presented as a replacement of $Fe^{2+}$ ion for $Fe^{3+}$ ion and for a negatively charged cation vacancy $^-$: $Fe^{2+} \to Fe^{3+}_{2/3} + ^-_{1/3}$.

Earlier we have observed [19] in the Mössbauer spectra the formation of a new SnO phase on the surface of $SnO_2$ crystallites in the process of oxidation of thin β-tin films. Formation of a surface layer of the tetragonal SnO phase is determined by the mechanism of β-tin oxidation. In the course of the oxidation process $Sn^{2+}$ ion leaves the crystal volume of $SnO_2$ and occupies the site on the surface because its ionic radius is longer than $Sn^{4+}$ ionic radius. Formally it can be presented as the transformation of a surface $Sn^{4+}$ ion of the $SnO_2$ rutile lattice in an $Sn^{2+}$ ion and in two surface anion vacancies $^+$: $Sn^{4+} = Sn^{2+} + 2\ ^+$, where $(Sn^{2+} + 2\ ^+)$ is an element of the defective SnO phase.

The same characteristics of all spectra recorded for the magnetite and the $SnO_2$ film samples after the PVMF is the increase of the intensity of spectral lines. Such spectral change of the same sample can be explained only by the increase of the Lamb-Mössbauer factor $f'$ for the sample after the PVMF. As the Lamb-Mössbauer factor $f'$ significantly depends on the degree of defectiveness of solids, the increase of the intensity of spectral lines after the PVMF shows the decrease of the degree of defectiveness of the sample.

Thus, the experimental results (Fig.1 and Fig.2) exhibit the decrease of the defectiveness of the magnetite structure after the PVMF and they are to be attributed, mainly, to the changes in the structure of the transitive layer (almost $\gamma$-$Fe_2O_3$) of magnetite. This can explain the MSE in $\gamma$-$Fe_2O_3$ [8]. The point is that both oxides ($\gamma$-$Fe_2O_3$ and $Fe_3O_4$) have the spinel structure and can be easily turned in each other. Consequently, the oxidation of $Fe_3O_4$ magnetite can lead to formation of the nonstoichiometric $\gamma$-$Fe_2O_3$ (which is quite probable). The effect of the magnetic field on the defective structure of this oxide can lead to its reduction to the initial $Fe_3O_4$ magnetite.

The change of $SnO_2$ after the PVMF (Fig.3) is to high extent relates to the transitive SnO phase, which either disappears, or significantly decreases. The increase of the intensity of the (110) reflection of $SnO_2$ rutile lattice in the X-ray spectra after the PVMF can be associated with the disappearance of transitive SnO phase and also with the reduction of the defectiveness in the surface structure of $SnO_2$ film.

Thus, it is natural to assume that the magnetic field directly affects the structural defects and promotes the process of their disappearance. The superstructural charged defects in the investigated oxide systems are the superstructural cation or anion vacancies. The superstructural charged defect in $Fe_3O_4$ was formed as a result of the capturing of $Fe^{2+}$ valent electron by the trap, which is the nearest cation vacancy . The return of the trapped electron from a cation vacancy $^-$ to $Fe^{3+}$ ion can lead to the disappearance of a superstructural defect in magnetite. Similarly, the return of the trapped holes from two anion vacancies $^+$ to $Sn^{2+}$ ion can lead to the disappearance of a superstructural defect in $SnO_2$.

Magnetic field promotes the return of a charge (electron or hole), captured by a trap (defect), to one of the nearest ions. It is known that the magnetic field can affect the spin dynamics in an electron-hole pair and promote their recombination. However their interaction is possible only when the electron and the hole come sufficiently close to each other. Until recently a possible mechanism of this process was described as thermal fluctuations [7], but we observed the MSE in the defective $Co_{3-x}O_4$ films [3] at temperatures of liquid helium and lower, at which thermal fluctuations are negligible. Therefore we suggested a mechanism, according to which a captured charge comes close enough to the center with an opposite charge. This mechanism is to be associated with the direct effect of the magnetic field on a charged dot defect.

In [4] the theoretical model of a low-energy electron trap (the charged dot defect) in a magnetic field is considered. An approximate analytic solution for the ground electron state is found to the Schrödinger equation for a combination of a uniform magnetic field and a single attractive delta-potential. It is known that the attractive delta-potential can be considered as the limit of narrow potential wells. The effect of the magnetic field on this loosely localized electron state is studied.

It was shown that this effect leads to appearing the probability current density in some region centered at the delta-potential as well as to changing the localization region of the electron in the plane perpendicular to the magnetic field. Both the factors promote an increase of probability of electron capture by positively charged center, which have got in area of electron localization. Loss of electron reduces stability of a trap in structure of a crystal and can lead to its destruction, i.e. to disappearance of defect.

The offered theoretical approach [4] does not exclude the contribution of the interaction of the magnetic field with defect spins in the MSE. However, in comparison with the direct effect of magnetic field on a charged dot defect, spin effects are considered to the effects of the second order.

The suggested theoretical model can be applied to variable (in time) magnetic fields too, if the magnetic field changes slowly (adiabatically) in comparison with characteristic times of the electron movement on quasiclassical orbit in a magnetic field and scattering of electron on one of positively charged centers brought into the area of electron localization.

Thus, in the experiments with the $Co_{3-x}O_4$ films [3] we used the constant magnetic field (H=3,5T) for the processing of samples. The direction of the field was perpendicular to the surface of the film. A very strong MSE was observed in $Co_{3-x}O_4$ films, and this can be explained according to the suggested model. In the experiments with the natural $Fe_3O_4$ magnetite and $SnO_2$ films the vortical magnetic field (H=0.1T, the frequency of 10 Hz) was used for the processing of samples. The MSE takes place in both studied systems. As the vortical magnetic field changes slowly (10 Hz) the suggested model can explain the observed MSE in these nonequilibrium oxides too.

**5. Conclusion.**
From these experiments it was concluded that in all studied nonequilibrium oxide systems MSE was the result of the direct interaction of the magnetic field with the charged dot defects of the crystalline structure. This can be explained in terms of the new microscopic model [4], which is the first step to create the general theory of the MSE. The theory should establish a complete chain of events initiated by a magnetic field at considerable microscopic and macroscopic levels in the nonequilibrium defective solids and should explain and predict the MSE in nonequilibrium defective solids. This theory is to provide the use of the processing in the magnetic field for control of modification of the defect structure and the properties of solids.

**Acknowledgements**
I am grateful to Dr. S. Kostyrev, Dr. A. Vasiljev and Dr. S. Reiman for the assistance in this work and also to Prof. V. Khalilov for fruitful discussion.

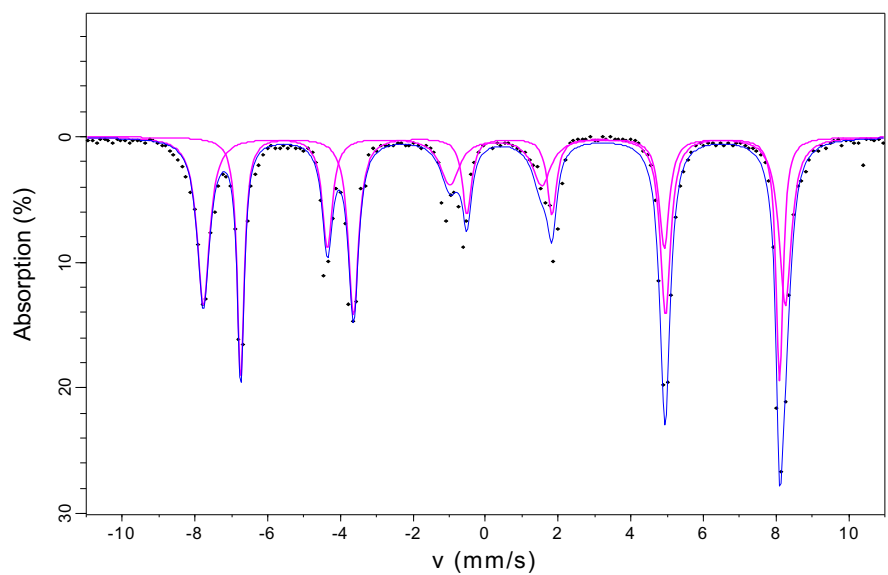

**Fig. 1(a)**
Typical Mössbauer spectrum measured for the $Fe_3O_4$ samples at 300K prior to the processing in magnetic field.

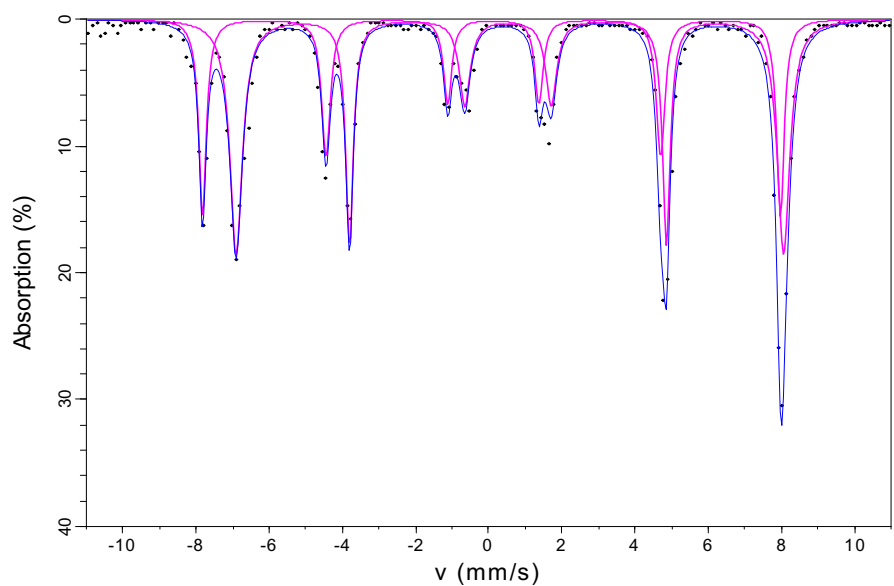

**Fig. 1(b)**
Typical Mössbauer spectrum measured for the $Fe_3O_4$ samples at 300K after the processing in vortical magnetic field with the intensity 0,1T and frequency of 10 Hz.

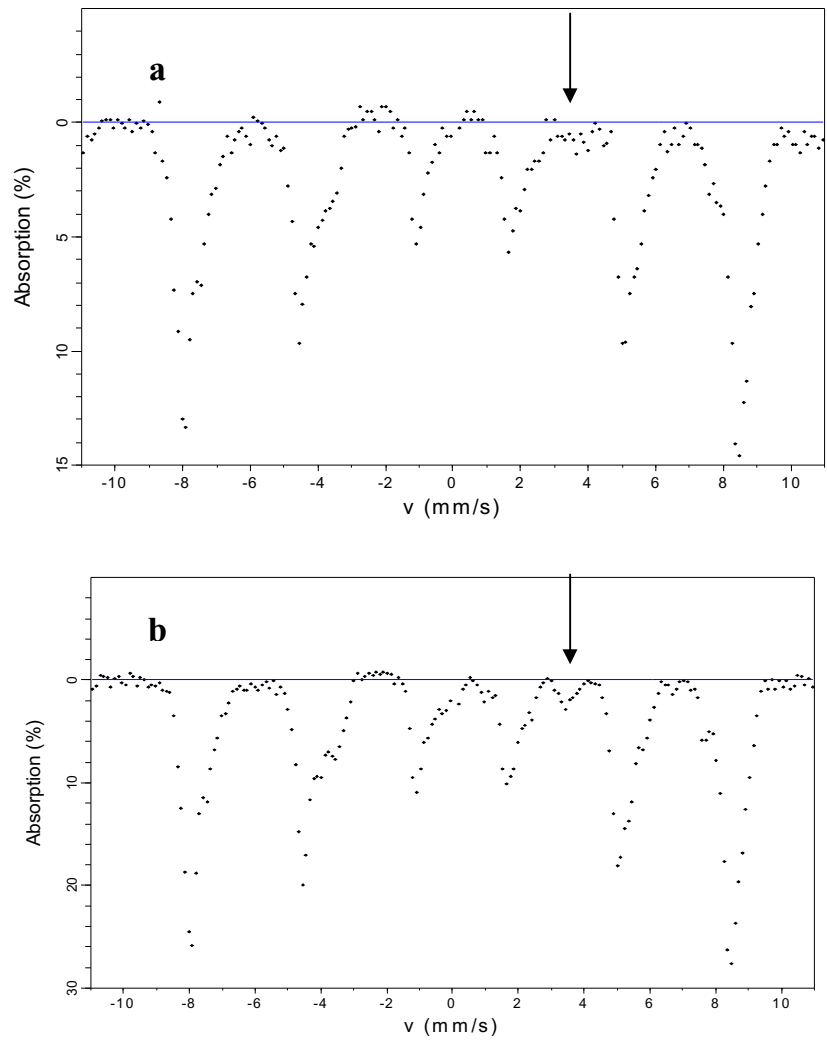

**Fig.2**
Typical Mössbauer spectra measured for the $Fe_3O_4$ samples at 77K:
(a) prior to the processing in a magnetic field;
(b) after the processing in vortical magnetic field with the intensity 0,1T, frequency of 10 Hz.

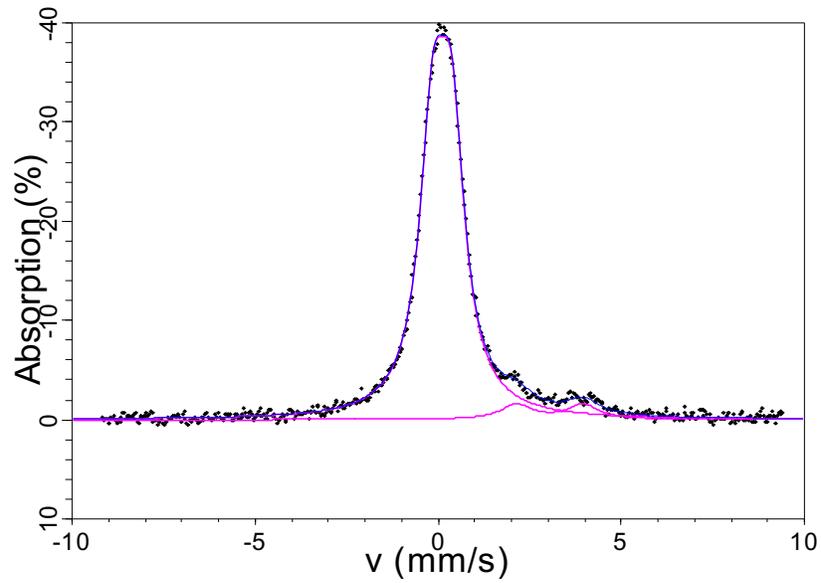

**Fig.3 (a)**
Typical Mössbauer spectrum measured for the $SnO_2$ film samples at 300K
prior to the processing in a magnetic field.

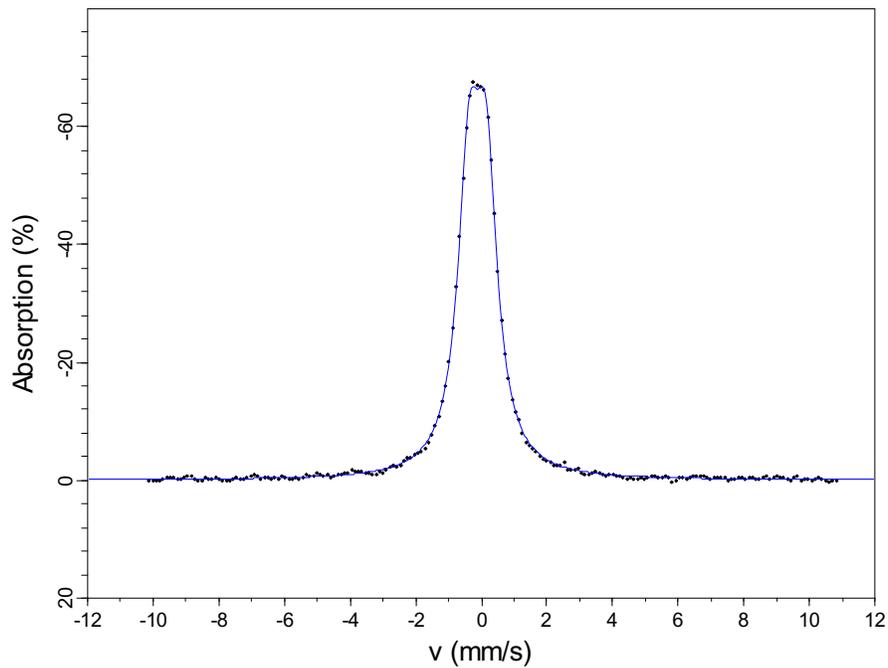

**Fig.3 (b)**
Typical Mössbauer spectrum measured for the $SnO_2$ film samples at 300K after the processing
in vortical magnetic field with the intensity 0,1T and frequency of 10 Hz.

**Table 1.**

Parameters of the Mössbauer spectra measured for the $SnO_2$ film samples prior to and after the processing in a vortical magnetic field with the intensity 0,1T and frequency of 10 Hz.

| Phase | δ (mm/s) | ΔE (mm/s) | Site Populations (%) |
|---|---|---|---|
| $SnO_2$ (prior to the processing) | 0.122±0.003 | 0.605±0.010 | 93.45 |
| SnO (prior to the processing) | 2.951±0.072 | 1.820±0.130 | 6.55 |
| $SnO_2$ (after the processing) | 0.081±0.003 | 0.551±0.010 | 100 |

**Table 2.**

X-ray data for the magnetite $Fe_3O_4$ prior to and after the processing in vortical magnetic field with the intensity 0,1T and frequency of 10 Hz.

| Sample (I, II – No of sample) | d, Å (hkl = 800) | a, Å (a=8d) | V, Å³ (V = a³) |
|---|---|---|---|
| $Fe_3O_4$ **(prior to the processing)** I II | 1.0508±0.0002 1.0506±0.0002 | 8.4064 8.4048 | 594.0 593.7 |
| $Fe_3O_4$ (after processing) I II | 1.0491±0.0002 1.0490±0.0002 | 8.3928 8.3920 | 591.1 591.0 |

**Table 3.**

X-ray data for the $SnO_2$ films prior to and after the processing in vortical magnetic field with the intensity 0,1T and frequency of 10 Hz.

| Sample (I, II – No of a sample) | a, Å | c, Å |
|---|---|---|
| $SnO_2$ **(initial)** I II | 4.741±0.002 4.743±0.002 | 3.195±0.002 3.193±0.002 |
| $SnO_2$ (after processing) I II | 4.741±0.002 4.741±0.002 | 3.193±0.002 3.192±0,.002 |